# Ethics-Based Auditing of Automated Decision-Making Systems: Intervention Points and Policy Implications

**Authors**:

Jakob Mökander[1]     ORCID: 0000-0002-8691-2582     Email: jakob.mokander@oii.ox.ac.uk

Maria Axente[2]     ORCID: 0000-0002-9488-3419     Email: maria_axente@yahoo.com

[1] Oxford Internet Institute, University of Oxford, 1 St Giles', Oxford, OX1 3JS
[2] Member of the Advisory Board, All-Party Parliamentary Group on AI (APPG AI), London, UK

Email for correspondence: jakob.mokander@oii.ox.ac.uk

## Abstract

Organisations increasingly use automated decision-making systems (ADMS) to inform decisions that affect humans and their environment. While the use of ADMS can improve the accuracy and efficiency of decision-making processes, it is also coupled with ethical challenges. Unfortunately, the governance mechanisms currently used to oversee human decision-making often fail when applied to ADMS. In previous work, we proposed that ethics-based auditing (EBA) – that is, a structured process by which ADMS are assessed for consistency with relevant principles or norms – can (a) help organisations verify claims about their ADMS and (b) provide decision-subjects with justifications for the outputs produced by ADMS. In this article, we outline the conditions under which EBA procedures can be feasible and effective in practice. First, we argue that EBA is best understood as a 'soft' yet 'formal' governance mechanism. This implies that the main responsibility of auditors should be to spark ethical deliberation at key intervention points throughout the software development process and ensure that there is sufficient documentation to respond to potential inquiries. Second, we frame ADMS as parts of larger sociotechnical systems to demonstrate that to be feasible and effective, EBA procedures must link to intervention points that span all levels of organisational governance and all phases of the software lifecycle. The main function of EBA should therefore be to inform, formalise, assess, and interlink existing governance structures. Finally, we discuss the policy implications of our findings. To support the emergence of feasible and effective EBA procedures, policymakers and regulators could provide standardised reporting formats, facilitate knowledge exchange, provide guidance on how to resolve normative tensions, and create an independent body to oversee EBA of ADMS.

## 1    Introduction: the rise of automated decision-making systems

Automated decision-making systems (ADMS)[1] increasingly permeate all aspects of society (AlgorithmWatch, 2019; Cath, 2018). Examples of the use of ADMS are abundant and include potentially sensitive areas like medical diagnostics (Grote & Berens, 2020) and the issuing of loans and credit cards (Lee et al., 2020). To perform tasks, ADMS rely on a plurality of statistical techniques, from decision trees to deep neural networks, to draw inferences from the growing availability of fine-grained data (Lepri et al., 2018). In this article, however, we will disregard the underlying technologies and instead focus on the features – e.g. autonomy, adaptability, and scalability – that underpin both socially beneficial and ethically problematic uses of ADMS.

On the one hand, delegating tasks to ADMS can bring both social and economic benefits. For example, it may lower costs, increase consistency, and enable new innovative solutions (Taddeo & Floridi, 2018). On the other hand, the use of ADMS is coupled with ethical challenges. ADMS may, for example, produce erroneous or discriminatory outcomes (Leslie, 2019). They can also enable human wrongdoing or remove human responsibility (Yang et al., 2018; Tsamados et al., 2020). Finally, a lack of trust in ADMS amongst citizens and consumers may cause significant social opportunity costs through the underuse of available technologies (Cookson, 2018). Taken together, the capacity to address the ethical risks posed by ADMS is quickly becoming a prerequisite for good governance (European Commission, 2019a).

Unfortunately, the tools for building ADMS have generally outpaced the growth and adoption of methods to understand whether such systems are reliable (Kroll et al., 2017; Schulam & Saria, 2019). Highly autonomous systems[2] may, for example, pose concerns associated with unintended impacts (Koene et al., 2019). Similarly, the ability of ADMS to update their internal decision-making logic over time makes it difficult to assign accountability when harm occurs (Burrell, 2016). Most pressingly, the delegation of tasks to ADMS curtails the sphere of ethical deliberation in decision-making processes (D'Agostino & Durante, 2018), as norms that used to be open for interpretation by human decision-makers are now embodied in ADMS.

To account for this shift, ADMS need to be designed and deployed in ways that are lawful, ethical, and technically robust (AI HLEG, 2019). Hence, to guide the design and use of ADMS, numerous governments, research institutes, companies and NGOs have published high-level ethics principles (Fjeld, 2020; Jobin et al., 2019).[3] These guidelines constitute a step in the right direction and tend to converge on very similar principles (Floridi & Cowls, 2019). However, the adoption and implementation of such guidelines remain voluntary. Moreover, when designing and operating ADMS, tensions may arise between different principles for which there are no fixed solutions (Kleinberg et al., 2017). Finally, the industry currently lacks both incentives and useful tools to translate abstract ethics principles into practical guidance on how to design and deploy ADMS (Morley et al., 2020). In short, there is a need to standardise and operationalise 'AI ethics' as a discipline (Kazim & Koshiyama, 2020b).

---

[1] With ADMS, we refer to autonomous self-learning systems that gather and process data to make qualitative judgements with little or no human intervention.

[2] Autonomous systems are often both complex and safety critical; this makes their formal specification and verification uniquely challenging (Luckcuck et al., 2019).

[3] Recent and reputable contributions include *Ethically Aligned Design* (IEEE, 2019), the *Ethics Guidelines for Trustworthy AI* (AI HLEG, 2019), and the *Recommendation of the Council on Artificial Intelligence* (OECD, 2019).

Against this backdrop, several recent publications have pointed towards auditing of ADMS as a promising governance mechanism to operationalise ethics (Bauer, 2017; Brundage et al., 2020; Kim, 2017). For example, Koshiyama et al. (2021) envision a new industry that focuses on assessing and assuring the legality, ethics, and safety of ADMS. Yet auditing of ADMS can be done in different ways and for different reasons (Bandy, 2021). For example, in their seminal paper *Auditing Algorithms*, Sandvig et al. (2014) proposed that audit studies should be used to investigate normatively significant instances of discrimination by ADMS operated by internet platforms. Such an approach would focus on the actual *impact* an ADMS has on its users and require the involvement of an *external auditor*. In contrast, Raji & Buolamwini (2019) suggested that *process-oriented, internal audits* can check that the engineering processes involved in the design and deployment of ADMS meet specific expectations or standards. All these approaches have merits, and different types of audits need not be mutually exclusive but can be crucially complementary.

Our focus here, however, is on so-called ethics-based auditing (EBA). Functionally, EBA is understood as a governance mechanism that helps organisations operationalise their ethical commitments (Mökander & Floridi, 2021). EBA thus concerns what ought and ought not to be done over and above the existing regulation. Operationally, EBA is characterised by a structured process whereby an entity's present or past behaviour is assessed for consistency with a predefined set of principles. Throughout this purpose-oriented process, various tools (such as software programs and standardised reporting formats) and methods (like stakeholder consultation or adversarial testing) are employed to verify claims and create traceable documentation. Naturally, different EBA procedures employ different tools and contain different steps. However, the main point is that EBA differs from simply publishing a code of conduct since its main activity consists of demonstrating adherence to a predefined baseline (ICO, 2020).

This is the second in a series of articles through which we consider the feasibility and efficacy of EBA as a governance mechanism that allows organisations to validate claims made about their ADMS. In our preceding work,[4] we defined what EBA *is*, how it *works*, and discussed *limitations* associated with the approach. We argued that EBA – if successfully implemented – can help organisations and societies mitigate the ethical challenges posed by ADMS through enhancing operational consistency and procedural transparency. While our analysis also suggested that EBA is subject to a range of conceptual, technical, economic, legal and institutional constraints, we nevertheless concluded that EBA should be considered as an integral component of multifaced approaches to managing the ethical risks posed by ADMS.

In this article, we shift focus from *what* EBA is and *why* it is needed – to *how* organisations can develop and implement effective EBA procedures in practice. The objective is twofold. First, we seek to identify the intervention points (in organisational governance as well as in the software development lifecycle) at which EBA can help inform ethical deliberation – and thereby make a positive difference to the ways in which ADMS are designed and deployed. Second, we seek to contribute to an understanding of how policymakers and regulators can facilitate and support the implementation of EBA procedures in organisations that develop ADMS. While we acknowledge that the development and implementation of EBA is an inherently practical activity, this article does not put forward a specific auditing procedure. Rather, it attempts to clarify the conditions under which EBA procedures can be feasible and effective.

In line with the objectives outlined above, we approach ethics from a governance perspective. This approach is based on the idea that proactivity in the design of ADMS, combined with process-oriented governance, can help identify risks and prevent harms before they occur (Kazim & Koshiyama, 2020a). Further, taking a governance perspective, we are compelled to view ADMS as part of larger sociotechnical systems (STS) that comprise people, organisations, and other technical artefacts. This implies that EBA procedures – to be feasible and effective – must consider not only the functionality, logic, and impact of ADMS (Dash et al., 2019) but also the entire ecosystem surrounding their use (Lauer, 2020).

Before proceeding, some clarifications will help narrow down the scope of this article. While the ethical challenges posed by ADMS are global in nature, this article aims primarily to support and inform the policy formation process within a European context. This geographical focus has some immediate implications. For example, we use 'ADMS' rather than the more popularised term 'artificial intelligence' (or 'AI-systems'). In doing so, we keep in line with the language used by AlgorithmWatch. However, 'AI' and 'ADMS' are often used interchangeably in the literature, so nothing hinges on this choice. Further, when discussing high-level ethics principles, we henceforth refer to the *Ethics Guidelines for Trustworthy AI* (AI HLEG, 2019). We do not argue that these ethical principles are better than those put forward by other institutions, but merely that adhering to the AI HLEG guidelines serves our purpose of investigating how EBA procedures can be developed within a coherent framework. This also implies that any comparison between, or evaluation of, different normative guidelines remains outside the scope of this article.

---

[4] The first article – on which the analysis and recommendations in this article builds – is titled 'Ethics-Based Auditing of Automated Decision-Making Systems: Nature, Scope, and Limitations' and was published in the journal *Science and Engineering Ethics* on 06 July 2021 (see Mökander et al., 2021).

Finally, we do not focus on any legal aspects of auditing. Instead, we conceptualise EBA as a soft, yet formal, 'post-compliance' governance mechanism (see Floridi, 2018). While hard governance refers to legally binding obligations that are precise and enforced, soft governance includes non-binding guidelines and norms (Erdelyi & Goldsmith, 2018). However, there is not conflict here: hard and soft governance mechanisms complement and reinforce each other. In this article, we therefore argue that the EBA is an essential complement to existing – and proposed – governance mechanisms, based on the assertion that decisions made by ADMS may be deserving of scrutiny even when they are not illegal.

The remainder of this article proceeds as follows. In section 2, we review previous work on EBA. In section 3, we frame ADMS as part of larger STS. In section 4, we analyse how STS are governed today. This enables us to identify intervention points at which ethical deliberation (informed by EBA) can help shape the behaviour of ADMS. In section 5, we acknowledge some important limitations of our approach. Finally, in section 6, we discuss the policy implications of our work and provide 8 recommendations on how policymakers can support the emergence of EBA as a feasible and effective governance mechanism that helps organisations address some of the ethical challenges posed by ADMS.

## 2   Previous work: introducing ethics-based auditing

In this section, we provide a brief overview of previous work in the field of EBA and summarise the most important findings from our preceding article in this series. Readers who are new to the field, or who seek an in-depth review of previous literature on EBA, are referred to our previous article (see Mökander et al., 2021). Before turning our attention to EBA specifically, however, something should be said about 'auditing' in the context of ADMS more generally.

As noted in the introduction, ADMS can both exacerbate existing inequalities (Eubanks, 2019; O'Neil, 2016) and introduce new types of ethical challenges (Coeckelbergh, 2020). For example, recent research suggests that ADMS may not only *discriminate* against specific individuals or groups (Datta et al., 2018; DeVries et al., 2019) but also *distort* information (Lurie & Mustafaraj, 2019; Robertson et al., 2018), *exploit* sensitive information without consent (Cabañas et al., n.d.; Vincent et al., 2019), and *misjudge*, i.e. make incorrect predictions or classifications (Bashir et al., 2019; Matthews et al., 2019). The aforementioned ethical challenges have given rise to a growing body of work that focuses on issues related to the fairness, accountability, and transparency of ADMS (Hoffmann et al., 2018). It has also led to the realisation that new governance mechanisms are needed to ensure that ADMS operate in ways that are legal, ethical, and technically robust.

Against this backdrop, 'auditing of algorithms' has recently attracted much attention from policymakers and academic researchers alike. For example, regulators like the ICO have drafted an auditing framework for AI (ICO, 2020; Kazim et al., 2021), and professional services firms like PwC (2019) and Deloitte (2020) are developing auditing (or 'assurance') frameworks to help clients verify claims made about their ADMS. Some of these initiatives focus on *compliance assurance*, i.e. by comparing a system to an existing set of standards, while other focus on *risk assurance*, i.e. the process of asking open-ended questions about how a system works (CDEI, 2021). EBA – as understood in this article – can refer to both compliance and risk-based approaches, as long as they are conducted voluntary and aim at demonstrating adherence to organisational values that exceed the baseline of legal permissibility.

Here, it should be noted that the idea of auditing software for consistency with predefined principles is not new. Since the 1970s, computer scientists have been involved in addressing issues of certifying software according to functionality and reliability (Weiss, 1980).[5] This means that more recent work builds on widely established and well-proven engineering and quality management processes. Koshiyama et al. (2021), for example, suggest that the activities that are (or should be) part of holistic auditing procedures include *development*, i.e. the process of developing and documenting the features of a specific ADMS; *assessment*, i.e. the process of evaluating the ADMS' behaviour and capacities; *mitigation*, i.e. the process of servicing or improving the outcome from an ADMS; and *assurance*, i.e. the process of declaring that an ADMS conforms to predetermined standards, practices or regulations.[6] For our purposes, the main takeaway here is that – to be feasible and effective – EBA procedures must link to intervention points that span all levels of organisational governance and all phases of the software lifecycle.

Outside the field of software development – and primarily in areas like finance – auditing has an even longer history of promoting trust and transparency (LaBrie & Steinke, 2019). Valuable lessons can be learned from these domains. Most importantly, the process of auditing is always purpose-oriented. Brown et al. (2021) have observed that different types of auditing could be used for different purposes, for example: (i) by regulators to assess whether a specific ADMS meets legal standards; (ii) by providers or end-users of ADMS to mitigate or control reputational risks; and (iii) by other stakeholders (including customers, investors, and civil rights groups) who want to make informed decisions about the way they engage with specific companies or products. While it is possible to design EBA procedures that address any of the purposes mentioned above, our focus in this article is on (ii) in Brown et al.'s typology. EBA is thus viewed as an organisational governance mechanism directed towards ensuring that ADMS operate in ways that align with specific ethics guidelines.

---

[5] See e.g. Rushby (1988) for an overview of the wide range of methods used at the time for testing and verifying the reliability of software systems from an engineering perspective.

[6] Note that while 'performance' standards presuppose known environments, more dynamic process-oriented standards can be applied throughout different phases of the software development lifecycle even when the environment is noisy or rapidly changing (Danks & London, 2017a).

Another lesson is that auditing presupposes operational independence between the auditor and the auditee. Whether the auditor is a government body, a third-party contractor, an industry association, or a specially designated function within larger organisations, the main point is to ensure that the auditing is run independently from the regular chain of command within organisations (Power, 1997). The reason for this is both to minimise the risk of collusion between auditors and auditees (IIA, 2017) and to be able to allocate responsibility for different types of system failures (more on this in section 4).

That said, let us now turn our attention back *ethics-based* auditing. Although widely accepted standards for EBA have yet to emerge, it is possible to distinguish between different approaches: *functionality audits*, for example, focuses on the rationale behind decisions. In contrast, *code audits* entail reviewing the source code of an algorithm. Finally, *impact auditing* investigates the types, severity, and prevalence of effects of an algorithm's outputs (Mittelstadt, 2016). These approaches are complementary and can be combined into holistic EBA procedures. However, since autonomous and self-learning ADMS may evolve and adapt over time as they interact with their environments, EBA needs to include (at least elements of) continuous, real-time monitoring (i.e., impact auditing).

In previous work (again, see Mökander et al., 2021), we reviewed the efficacy and feasibility of EBA of ADMS. Our analysis suggested that, as a governance mechanism, EBA displays several methodological advantages. For example, EBA can help allocate accountability for potential system failures, e.g. by demanding that design decisions made throughout the software development process are documented and that outputs from ADMS are monitored throughout the systems' lifecycle. In this respect, tools like model cards, i.e. documents that provide basic information about the properties and limitations of ADMS (Mitchell et al., 2019; Saleiro et al., 2019), are useful.

In addition, EBA can provide decision-making support to software developers and policymakers alike, e.g. by leveraging tools that help visualise the normative values embedded in ADMS (AIEIG, 2020). One such tool is FAIRVIS, which offers a visual analytics system that integrates a subgroup discovery technique to scrutinise ADMS in order to inform normative discussions around group fairness (Cabrera et al., 2019). Other examples include the TuringBox (Epstein et al., 2018) and the What-If-Tool (Google, 2020), both of which allow software developers to test the performance of their ADMS in hypothetical situations. Such insights help inform ethical deliberation at critical points in the process of developing ADMS.[7]

Yet, the benefits of EBA are potential and not guaranteed. The extent to which they can be realised in practice depends on the design of auditing frameworks and complex environmental factors. Moreover, EBA of ADMS is subject to a range of conceptual, technical, economic, and institutional constraints. Three of them are worth highlighting here.

First, EBA is constrained by the difficulty of quantifying externalities that occur due to indirect causal chains over time (Rahwan, 2018). For example, while practitioners are encouraged to consider – and account for – the social implications of a prospective ADMS throughout the software development process, this is often difficult in practice since scalable and autonomous systems may have indirect impacts that spill over borders and generations (Dafoe, 2018). Hence, rather than attempting to codify ethics, one function of auditing is to arrive at resolutions that, even when imperfect, are at least publicly defensible (Whittlestone et al., 2019).[8]

Second, a weakness of traditional auditing methodologies is the assumption that test environments sufficiently mimic the later application to allow quality assurance (Auer & Felderer, 2018). Put differently; there is a tension between the stochastic nature of ADMS and the linear, deterministic nature of conventional auditing procedures. As a result, the same agile qualities that help software developers to meet rapidly changing customer requirements also make it difficult to ensure compliance with pre-specified requirements (Steghöfer et al., 2019). This implies that EBA must monitor and evaluate not only performance-based criteria but also process-based criteria.

Third, from a social perspective, there is always the potential for adversarial behaviour during audits. The organisation or ADMS that is being audited may, for example, attempt to trick the auditor by withholding information or temporarily adjusting its behaviour. While many auditing frameworks to some extent anticipate adversarial behaviour, so-called 'management fraud' can still evade auditors. Similarly, even if audits reveal flaws within ADMS, asymmetries of power may prevent corrective steps from being taken (Kroll, 2018).

Bearing the aforementioned constraints in mind, it is clear that EBA is not a blanket solution for all ethical challenges posed by ADMS. Nevertheless, as a soft yet formal governance mechanism (more on this in section 4), EBA can help organisations that genuinely seek to adhere to specific ethics guidelines when designing and deploying ADMS to do so more effectively.

---

[7] In this regard, EBA is akin to what Floridi (2016a) calls informational nudging: it changes the nature of the information to which an agent is exposed in order to make a decision or obtain a goal.

[8] This approach mirrors the claim made by Gabriel (2020) that the main challenge is not to identify 'true' moral principles but rather to identify procedures that receive endorsement despite widespread variation in people's beliefs.

However, in order to consider *how* we must first take a step back to analyse the role ADMS play in larger STS.

## 3    A systems approach: decision-making in complex environments

In the broadest sense of the term, a system is the ordering of a set of parts into a whole. An STS is one that comprises both social entities, like people and organisations, and technical entities, like tools, infrastructures, and processes (Chopra & Singh, 2018). ADMS, then, refers to technical systems that encompass decision-making models, algorithms that translate models into computable code, as well as methods to acquire and process input data. ADMS interact with the entire political and economic environment surrounding their use. Figure 1 below illustrates how ADMS are situated within larger STS.

*Figure 1. ADMS are part of complex sociotechnical systems (STS)*

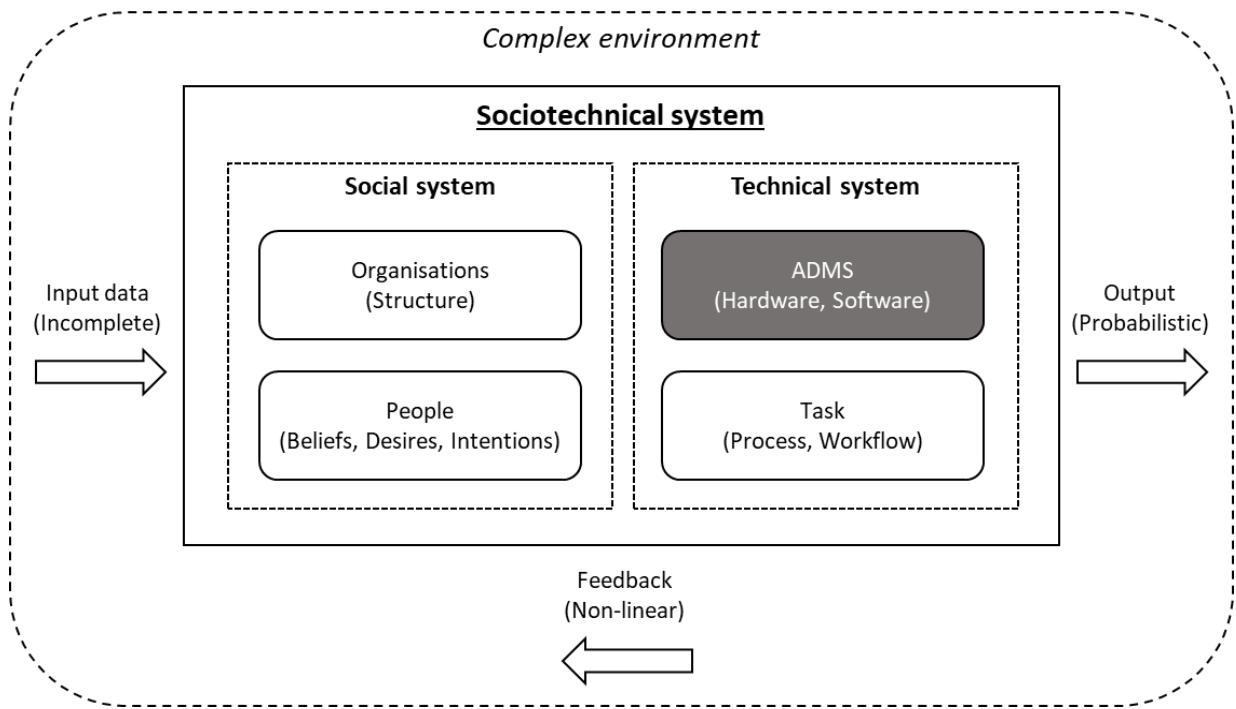

As figure 1 shows, ADMS must be integrated into broader STS to work (Eubanks, 2019). However, the different components of an STS are not always easy to distinguish since their relations are dynamic and subject to change over time (Di Maio, 2014). Moreover, a given system can be described at various levels of abstraction (LoA), and the appropriate LoA depends less on the features of the system under consideration than on the purpose of the analysis (Floridi, 2008). Because we are interested in investigating the merits of EBA of ADMS, we here take the task, i.e., the decision-making task itself, as the starting point for our analysis.

Decisions that impact humans and their environments are typically performed within specialised organisations, as when, for example, a healthcare provider diagnoses a patient for suspected skin cancer or a social media platform customises the content it displays. In each case, the decision-making task not only demands specific input data but also determines what constitutes a good outcome. To continue with our examples, while accuracy may be the most important principle when diagnosing skin cancer, other considerations like data protection may reasonably limit the extent to which social media platforms can customise ad deliveries. The point here is not to discuss what constitutes an appropriate balance between different values but rather to highlight that whether it is ethically justifiable to treat an individual in a specific way is a nuanced moral question that is likely to depend on multiple contextual factors.

Ultimately, the decisions under investigation are made by people (e.g., doctors or software developers) rather than by the organisations they represent. However, the structure, culture, interests of different stakeholders, and incentives of organisations affect the decisions made by the people who work in them (Hodges, 2015), and all tools available to human experts in the decision-making process fulfil and have been designed for, specific functions (Van de Poel, 2020). Because they alter the cost-benefit ratio of different potential actions, technical artefacts inevitably influence human decision-making (Danaher, 2017) and should not be viewed as value-neutral (Floridi, 2017a). However, in this regard, ADMS are no different from other tools that support human decision-making. Rather, the point to be stressed here is that ADMS and their environments, or 'ecosystems' (Lauer, 2020), co-evolve in ways that are both nonlinear and dynamic, through which ADMS may acquire features as they interact with their environments. It follows that the link between a decision and the impact this decision has on its environment is not always intuitive, nor need it be consistent over time.

To deal with this complexity, various components within STS exchange information through feedback loops. In fact, feedback is the central mechanism that allows functions to be performed correctly and enables goals to be achieved (Di Maio, 2014). Because human experts are reflective, they will (often unconsciously) monitor and evaluate the outcomes and compare them with their own values as well as with the values embodied in the system (Van de Poel, 2020). In contrast, ADMS promote technological normativity and thus tend to curtail the space for social reflexivity (i.e. the public discussion and evaluation of norms) in the decision-making process (D'Agostino & Durante, 2018).

That said, both human decision-makers and ADMS come with their own sets of strengths and weaknesses (Baum, 2017). For instance, human judgement may be influenced by prejudices, fatigue or hunger (Kahneman, 2012). The use of ADMS can, therefore, in some cases, lead to more objective and potentially fairer decisions than those made by humans (Lepri et al., 2018). Moreover,

ADMS may enhance the overall efficiency of decision-making processes by optimising bureaucratic procedures and accessing real-time data. This has led researchers like Zerilli et al. (2018) to argue that ADMS are being held to unrealistically high standards compared to the available alternatives. Such objections are not necessarily wrong. But comparable performance with human decision-makers is only a necessary condition for ethically-sound automated decision-making, not a sufficient one. Over and above such a minimal criterion, superior performance by ADMS may be needed to offset potentially unwelcome side-effects as well as seen and unforeseen threats (Tasioulas, 2018).

What are the implications of considering ADMS from a system perspective for the prospects of subjecting them to EBA? As we have seen, important dynamics of the STS as a whole may be lost or misunderstood if either the social or the technical subsystems are targeted separately (Di Maio, 2014). This risk is summarised by what Lauer (2020) calls the fallacy of the broken part: when there is a malfunction, the first instinct is to identify and fix the broken part. In practice, however, most serious errors or accidents associated with ADMS can be traced not to coding errors but to requirement flaws (Leveson, 2011).[9] This means that no purely technical solution will be able to ensure that ADMS behave in ways that are ethically sound.[10]

From an auditing perspective, viewing both organisations and ADMS through a systems LoA has several advantages. First, a system is characterised not only by its parts but also by the relations among them (Hanneman, 1988). This allows EBA procedures to shift focus from the analysis of static entities to the evaluation of dynamic processes. Second, a system can always be understood in terms of its input and outputs (Kroll, 2018). This gives a theoretical foundation for functionality audits, whereby ADMS can be evaluated based on their intended and operational goals. With intended goals, we refer to those of the human designers, owners, and users of ADMS. As Samuel (1960) put it, the machine does not possess a will, and its conclusions are only the logical consequences of its input. Third, given a complex environment in which system outputs often have nonlinear impacts, we can conclude that EBA of ADMS needs to consider not only the system's source code and the purpose for which it is employed but also the actual impact it exerts on its environment over time as well as the normative goals of relevant stakeholders.

---

[9] Requirement flaws include erroneous or inadequate task description, operational procedures or lacking testing requirements. The most common requirement flaw is incompleteness. See Leveson (2011) for further details.

[10] The 'solutionism trap' occurs when it is assumed that technical solutions alone can solve complex social and political problems. According to Andrus et al. (2021), a first step to avoid the solutionism trap is to maintain a robust culture of questioning which problems should be addressed, and why. A second step would be to examine which properties are not tied to the technical objects under investigation but to their social contexts.

## 4 Governing sociotechnical systems: identifying intervention points for EBA

In this section, we analyse how complex STS are governed today and discuss how EBA procedures can be designed to complement and enhance existing governance structures. Governance is an interactive activity among stakeholders whereby they try to align the norms of an STS with their values (Chopra & Singh, 2018) by exerting power, authority and influence and enacting policies (Baldwin & Cave, 1999). Thus governance consists of both hard and soft aspects.

Hard governance mechanisms are systems of rules elaborated and enforced through institutions to govern the behaviour of agents (Floridi, 2018). When considering ADMS, examples of hard governance mechanisms range from legal restrictions on system outputs to outright prohibition of the use of ADMS for specific applications (Koene et al., 2019). Soft governance embodies mechanisms that abide by the prescriptions of hard governance while exhibiting some degree of contextual flexibility, like subsidies and taxes. Although the ethics guidelines published by AI HLEG are an example of soft governance, they presuppose and are aligned with the EU legislation (Floridi, 2019a). However, from an ethical perspective, there will always be situations where it is better not to do something, even if it is legal, and, inversely, situations where it is better to do something even if it is not legally required (Floridi, 2019b).

Consider, for example, the emerging regulatory landscape in the EU. On 21 April 2021, the European Commission (2021) published its proposal for a new Artificial Intelligence Act (AIA).[11] According to the AIA, some use cases of ADMS will be completely banned,[12] whereas other 'high-risk' systems will be subjected to mandatory 'conformity assessments'[13] and 'post-market monitoring'.[14] These enforcement mechanisms proposed in the AIA are examples of hard governance.[15] However, the AIA still leaves room for post compliance, *ethics-based* auditing (Mökander et al., Forthcoming). Most notably, providers of non-high-risk ADMS are encouraged to draw up and apply voluntary codes of conduct (AIA: Article 69) related to their internal procedures and the technical characteristics of the systems they design and deploy. Further, even providers of high-risk ADMS may benefit from adopting voluntary codes of conduct that go over and above the requirements set out in the AIA, e.g. to manage reputational risk or to position their services to specific audiences (Buhmann et al., 2020). In both these cases, EBA can complement the AIA and help organisations verify the claims made about their ADMS. Therefore, to foster the development of ethical ADMS, it is necessary to complement hard governance mechanisms with soft mechanisms, including economic incentives and increased public deliberation (Binns, 2018).

---

[11] The AIA is the first attempt to elaborate a general legal framework for 'AI' carried out by any major global economy.
[12] This includes the prohibition of ADMS used for general-purpose social scoring and real-time remote biometric identification of natural persons in public spaces for law enforcement (AIA's Explanatory Memorandum: p. 21).
[13] Through such conformity assessments, providers can show that their high-risk systems comply with the requirements set out in the AIA ex-ante, or 'before placing the system on the market.
[14] Post-market monitoring' refers to the requirement that providers must document and analyse the performance of high-risk ADMS throughout their lifetimes (AIA: Article 61).
[15] According to the AIA, organisations that supply incorrect or incomplete information may be subject to fines ranging up to 10,000,000 EUR (or 2% of global annual turnover).

A further distinction can be made between formal and informal governance mechanisms. Formal governance mechanisms are officially stated, communicated and enforced (Falkenberg & Herremans, 1995). While hard governance mechanisms are formal per definition, not all formal governance mechanisms are necessarily hard. Budgets, codes of conduct and reward criteria are, for example, soft yet formal governance mechanisms. Informal governance, on the other hand, comprises common values, beliefs and traditions that direct the behaviour of individuals and groups within STS. Since people tend to follow rules where they correspond to their internal moral value system (Hodges, 2015), organisational culture is a strong determinant of ethical behaviour.[16]

The governance of complex STS will be most effective when a wide range of complementary mechanisms pull in the same direction (Hodges, 2015). To complement and strengthen the congruence of existing governance structures within organisations that develop and use ADMS, EBA should, therefore, be thought of as a soft yet formal governance mechanism. Rather than imposing prescriptive, top-down solutions, the affordances of EBA are well suited to both to provide regulators and decision-subjects with justifications for the outputs produced by ADMS and inform ethical deliberation within organisations at critical intervention points that shape ADMS and their behaviour. The former claim is supported by the theory of public reason. From this perspective, just as democratic citizens have a right to scrutinise and hold accountable the exercise of political power, decision-subjects should have the right to scrutinise and hold accountable the exercise of ADMS (Binns, 2018). The latter claim draws on standard practices within business ethics: the behaviour of an organisation is embedded in its operating model, i.e., through processes, roles and responsibilities, incentives, policies, etc., and hence inevitably linked to governance as a whole (Crane & Matten, 2016). This is because line managers in organisations that design ADMS have a responsibility to work towards the achievement of organisational objectives – including managing risks and ensuring adherence to organisational values.

As illustrated by Figure 2 below, potential intervention points (i.e., points at which decisions, actions, or activities are likely to shape the design and behaviour of ADMS) span all levels of organisational governance and all phases of the software lifecycle. First, at an organisational level, intervention points include employee training and incentive structures. These are particularly important since fundamentally unethical organisations would not have the capabilities to build and deploy ethical ADMS (Lauer, 2020). Further, software developers can and should engage in ethical deliberation throughout the product development process (Chopra & Singh, 2018) by considering questions such as: does the ADMS satisfy the values of relevant stakeholders? Has relevant feedback regarding outputs and system impacts been taken into account when operating and redesigning ADMS? And do social actors within STS behave in ways that are socially acceptable? The core idea here is that influence over the ethical impact that ADMS have can be exerted at the time of their design (Turilli, 2008).

---

[16] Indeed, informal governance mechanisms are often the dominant source of control in resolving ethical issues (Falkenberg & Herremans, 1995).

*Figure 2. Examples of intervention points throughout the ADMS lifecycle*

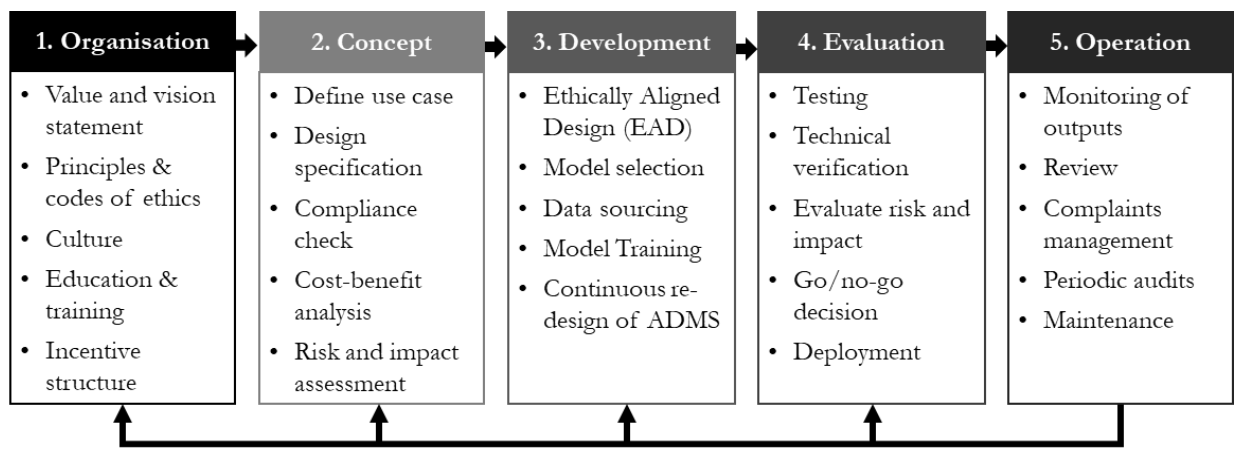

Responsibility for identifying and executing relevant interventions lies with the management of organisations that design and deploy ADMS. However, EBA can help inform ethical deliberations at different intervention points insofar as the independent auditor works together with the auditee to ensure that the right questions have been asked (and adequately answered). The auditor should also assess and verify the claims made by the auditee about its processes and products. Finally, the independent auditor should ensure that there is sufficient documentation and empirical evidence to respond to potential inquiries from public authorities or individual decision-subjects.

But which are the relevant questions that should be asked?[17] And how much documentation is sufficient? These are difficult questions that can only be answered contextually, given a specific decision-making scenario and context. Similarly, since there are correlations and interdependencies between all activities within STS and the resultant behaviour of ADMS, any list of potential intervention points would be inexhaustive. Nevertheless, we have attempted to highlight some of the intervention points at which EBA can help shape the design and deployment of ethical ADMS by informing ethical deliberation. These are displayed in Table 1 below.

---

[17] Here, a parallel can be made to Dean et al. (2021), who presented a framework for sociotechnical inquiry along axes of *value*, *optimization*, *consensus*, and *failure* as an initial toolbox to gauge a systems' relationship with society.

*Table 1. Intervention points at which EBA can help inform ethical deliberation*

| Intervention point | EBA should inform and document ethical deliberation about: |
|---|---|
| 1) Value and vision statement | Whether value and vision statements are publicly communicated<br>How the behaviour of ADMS reflects value and vision statements |
| 2) Principles and codes of conduct | How principles and codes of conduct are translated into organisational practices, i.e., provide practical guidance |
| 3) Ethics boards and review committees | What are the pathways through which ethical issues can be escalated, tensions can be managed, and precedents can be set |
| 4) Stakeholder consultation | What is the perceived impact of ADMS on decision-subjects and their environment |
| 5) Employee education and training | Whether ethical considerations are regarded in training programmes<br>What tools and methods do employees have at their disposal to aid ethical thinking, analysis and reasoning |
| 6) Performance criteria & incentives | What types of behaviour do existing reward structures incentivise<br>How well performance criteria support stated values and visions |
| 7) Reporting channels | How to provide avenues for whistleblowing that enables organisational learning whilst avoiding a 'culture of blame' |
| 8) Product development | Which tradeoffs have been made in the design phase and why<br>What ethical risks are associated with intended and unintended uses of the ADMS |
| 9) Product deployment and redesign | Whether and how the ADMS has been piloted prior to deployment<br>How continuous monitoring and stakeholder consultation can inform the continuous redesign of the ADMS |
| 10) Periodic audits | Whether periodic audits account for and review the ethical behaviour of organisations and ADMS<br>How decisions and processes are documented, stored, and communicated for transparency and traceability |
| 1) Monitoring of outputs | Are there adequate feedback mechanisms that alert or intervene if decisions made by ADMS transgress a predefined tolerance span |

Many of the intervention points listed above already exist within organisations that design and deploy ADMS. Thus, implementing EBA does not necessarily require any additional layers of governance to be imposed top-down. Rather, what EBA should do is inform, formalise, assess, and interlink existing governance structures, thereby complementing and enhancing them. It would be beyond the scope of this article to describe the complexities associated with informing ethical deliberation at each potential intervention point. Nevertheless, we next discuss three examples in order to highlight the mechanisms whereby EBA helps translate organisational values into ethical behaviour in ADMS.

Let us first consider how principles and codes of conduct are used within organisations. As stand-alone documents, they do little to bring about real change in how individuals behave (McNamara et al., 2018). Yet regulatory mechanisms are most effective when adopted transparently and consistently (Hodges, 2015) with executive-level support (Floridi & Strait, 2020). Hence, by formalising and legitimising the use of codes of conduct throughout the organisation, EBA provides a basis to hold individual agents accountable for their actions (or lack of) and to improve the impact that the use of codes of conduct has on organisational decisions in the first place.

Next, consider the role of incentives and reward structures. Human beings are not perfectly rational agents, and unethical behaviours are not solely caused by conflicting incentives. Individual character traits like idealism and job satisfaction also play important roles (Hagendorff, 2020). However, if the reward structures for people involved in designing and deploying ADMS are misaligned with the stated values and vision of the organisation, we have little reason to expect that the resulting behaviour of the ADMS they produce will be ethically aligned (Brundage et al., 2020). Moreover, agents are most likely to obey norms when they perceive the risk of being identified as high.[18] Building on these insights, EBA helps rectify skewed incentive structures in two ways. First, when conducted by an independent third-party, EBA can help ensure that organisational practices are subjected to external scrutiny. Thereby, the focus of the auditor should be to assess and report on the alignment between actual incentive structures and official value and vision statements. Second, from an internal perspective, EBA changes the incentive structure by reinforcing cultural norms for what constitutes acceptable conduct, thereby reducing the risk for neglect or malicious behaviour.

Finally, let us spell out how continuous monitoring of system outputs helps inform ethical deliberation amongst agents in STS and, by extension, contributes to the (re)design of ADMS. As discussed in section 3, different components of STS influence one another and their environment through feedback loops. This means that there are no simple answers about how different micro-actions aggregate to the system level (Lauer, 2020). Feedback provides pragmatic ways of dealing with this complexity: by shifting the focus from describing the system to designing a blueprint for what success looks like, we can participate proactively in shaping future societies (Floridi, 2017b). In the case of designing ADMS, both social and technical feedback loops are necessary. Social feedback channels provide decision-subjects with avenues for questions and appeals. They also create an opportunity for system owners to provide relevant and intelligible explanations (PwC, 2019). Technical feedback can be used to identify and learn from the root causes of problematic behaviour, test the validity of an ADMS's logic and answer critical questions like to what extent the undesired behaviour derives from the data, the design of the ADMS itself, or from how the results are used (Bauer, 2017). The basic idea is that sustained and continued assessment ensures that continuous feedback regarding emerging ethical concerns from an ADMS is worked into the next iteration of its design and application (Tran & Daim, 2008).

---

[18] Here, we are concerned with 'informal' norms. For an extensive discussion of the role of formalised, professional norms in 'AI governance', please see Gasser & Schmitt (2019).

## 5    Discussion: limitations and risks associated with EBA

The extent to which EBA can contribute to ensuring that ADMS behave ethically depend not only on how auditing procedures are designed but also on the intent of stakeholders. An analogy borrowed from Floridi (2014) is helpful to illustrate this point: the best pipes may improve the flow but do not improve the quality of the water, yet water of the highest quality is wasted if the pipes are rusty or leaky. Like the pipes in the analogy, EBA is not morally good in itself, but it enables moral goodness to be realised if properly designed and combined with the right values. EBA is an *infraethical* element of STS, it facilitates the delivery of ethically sound outcomes, but it is not per se sufficient to ensure such outcomes (Floridi, 2014).

That being said, even the best efforts to translate ethics principles into organisational practices may be undermined by a set of ethical risks. Floridi (2019) lists five such risks. For our purposes, the three most relevant of these risks are: *ethics shopping*, i.e., the malpractice of cherry-picking ethics principles to justify pre-existing behaviours; *ethics bluewashing*, i.e., the malpractice of making unsubstantiated claims about the ethical behaviour of an organisation or an ADMS; and, *ethics lobbying*, i.e., the malpractice of exploiting 'self-governance' to delay or avoid necessary legislation about the design of ADMS.

EBA is not immune to these concerns. For instance, consider the keen interest taken by large technology companies in developing tools and methods for EBA. While commendable, experiences from self-governance initiatives in other sectors suggest that the industry may not want to reveal insider knowledge to regulators but instead use its informational advantage to obtain weaker standards (Koene et al., 2019; Selbst, 2021). Yet the fact that EBA does not resolve all tensions associated with the governance of STS is not necessarily a failure. Rather, it is to the credit of EBA that it helps manage some of these tensions. For example, by demanding that ethics principles and codes of conduct are clearly stated and publicly communicated, EBA ensures that organisational practices are subject to additional scrutiny, which, in turn, may counteract ethics shopping. Similarly, when conducted by an independent auditor and provided that the results are publicly communicated, EBA can also help reduce the risk of ethics bluewashing by allowing organisations to validate the claims made about their ethical conduct and the ADMS they operate.

While good governance is about balancing conflicting interests, it can take time for socially acceptable equilibria to form. Hence, new types of governance mechanisms often suffer from shortcomings like *tunnel vision* – whereby overregulation may end up doing more harm than good; *random agenda selection* – whereby priorities are set by special interest groups; and *inconsistency* – whereby different standards are used to evaluate different options (Baldwin & Cave, 1999). As a soft yet formal governance mechanism, EBA may suffer from these limitations.

First, let us consider the risk of tunnel vision. It is true that decisions based on incomplete or biased data may end up being erroneous or discriminatory and that the abilities of ADMS to draw non-intuitive inferences may infringe on privacy rights. Nevertheless, when governing new technologies like ADMS, we must be careful not to optimise a single value at the expense of others (Muhlenbach, 2020). Here, the use of multiple evaluation metrics and tolerance intervals can help improve the comprehensiveness of the ethical evaluation, thereby minimising the risk of tunnel vision (Suresh & Guttag, 2019).

Second, normative values often conflict and require tradeoffs. For example, ADMS may improve the overall accuracy but discriminate against specific subgroups in the population (Whittlestone et al., 2019). Similarly, different definitions of fairness – like individual fairness, demographic parity and equality of opportunity – are mutually exclusive (Kusner et al., 2017). Because fundamental political disagreements remain hidden in normative concepts, the development of EBA methodologies runs the risk of random agenda selection, whereby EBA are designed with specific, yet partial or unjustified, normative visions.

Finally, it would be unrealistic to expect decisions made by ADMS to be any less complicated to evaluate from an ethical perspective than those made by humans.[19] Ethical decision-making inevitably requires a frame of reference, i.e., a baseline against which normative judgements can be made. If analysed in a vacuum, ADMS risk being held to higher standards than available alternatives (Zerilli et al., 2018). Such inconsistencies may, in some cases, end up doing more harm than good, as when, for instance, a particular ADMS is not used due to concern about accuracy or bias – even if it performs better than humans on the same measures. This absolutism ties back to the naïve belief that we have to – or indeed even can – resolve disagreements in moral philosophy before we start to design and deploy ADMS. However, a more nuanced approach would be to understand ADMS in their specific contexts and compare them to the relative strengths and limitations of human decision-makers (Whittlestone, Nyrup, et al., 2019).

It should also be noted that when analysing STS, there is an inevitable three-way tradeoff between brevity, depth, and breadth. The systems approach on which our analysis is based prioritises breadth. This is not unproblematic yet can be motivated by the need to understand how EBA relate to existing organisational governance structures. However, a systems approach inevitably leaves unanswered many questions at more detailed LoA. To mention a few: was the right ADMS built, and was it build right? (Dobbe et al., 2019); what are the appropriate ethical assessment criteria for ADMS? (D'Agostino & Durante, 2018); what should be included when documenting the origin of a dataset or the design of an ADMS? (Raji et al., 2020); how to account for the power asymmetries between system owners, regulating bodies and decision-subjects when designing EBA procedures? (Crawford et al., 2019); and, who within STS is responsible for distributed moral action? (Floridi, 2016b). These questions are left for future research. Nevertheless, already at the high LoA adopted in this article, a number of policy implications become clear. In the next section, we will discuss how policymakers and regulators can facilitate the adoption of EBA by organisations that design and deploy ADMS.

---

[19] Some problems are simply 'hard' because human values are not always amenable to quantitative representation. See e.g. Goodman (2021) for an excellent discussion on the limits to rational decision-making.

## 6    Implications: policy recommendations

Initiatives to develop EBA processes taken by different organisations – including private companies, academic researchers, industry associations and consumer protection and human rights groups – should be encouraged. In fact, organisations that design and deploy ADMS have good reasons to subject themselves and the systems they operate to EBA. For example, ensuring the ethical alignment of ADMS would help organisations manage financial and legal risks (Koene et al., 2019). Moreover, it may also help them gain competitive advantages (European Commission, 2019b): just as organisations seek to certify that their operations are sustainable from an environmental point of view (IEEE, 2019), or demonstrate to consumers that their products are healthy through detailed nutritional labels (Holland et al., 2018), the documentation and communication of the steps taken to ensure that ADMS are ethical can play a positive role in both marketing and public relations.[20]

To achieve these potential benefits, much can and should be done voluntarily at an organisational level. Amongst others, and in line with best practices for business ethics (Hodges, 2015), providers and operators of ADMS ought to: consult with all staff on what values the firm stands for; publish clear statements about the values and principles they seek to adhere to; obtain external inputs and utilise that feedback in the continuous redesign of ADMS; align internal reward structures with the stated organisational values; create a culture that encourages learning and allows employees to speak up; and, create means for demonstrating the extent to which they – and the ADMS they operate – adhere to their ethical values. This would require ethically aligned upskilling, as well as active organisational change management via employee engagements.

Despite the incentives for organisations to voluntarily adopt and implement EBA procedures, industry self-regulation is coupled with inherent challenges (as discussed in section 5 above). However, act-stage interventions (like EBA) may still be preferable to harm-stage control when it is difficult to hold firms or individuals to account for causing harm (Baldwin & Cave, 1999).

---

[20] Independent assessments of ADMS can help organisations improve a number of business metrics, including talent acquisition, regulatory preparedness, data security, reputational management, and process optimisation (EIU, 2020).

Especially in a fast-changing field like the development and deployment of ADMS, and insofar as the risk of ethics lobbying can be curtailed, self-governance based on soft yet formal mechanisms can provide a valuable additional 'layer of governance' (Gasser & Almeida, 2017) – complementary to that of existing and future legislation.

Further, how well industry self-regulation works as part of a multifaced approach to a specific problem is not independent of the guidance and support provided by policymakers (OECD, 2015). In fact, policymakers and regulators can do much to facilitate the emergence, incentivise the adoption, and strengthen the effectiveness of different EBA procedures. Here, we highlight eight policy recommendations that follow from our analysis in this article. Specifically, policymakers and regulators should consider to:

1. Help provide working definitions for ADMS

   Different types of ADMS pose different ethical challenges (Adler et al., 2018). Even for systems that are technically similar, the ethical risks they pose may vary across different use cases, both within and between sections (AIEIG, 2020). Moreover, both the level of autonomy and the ability to learn displayed by ADMS are matters of degree. As a result, it can often be difficult for organisations to understand (and identify) which of their systems ought to be subjected to EBA. However, every policy needs to define its material scope (Schuett, 2019). Regulators can help organisations demarcate the material scope for EBA by providing working definitions (or risk classifications) of ADMS that enable proportionate and progressive governance. Efforts in this direction have already been initiated (see, e.g., OECD's (2020) *draft framework for the classification of AI systems*). The creation of working definitions of ADMS – as well as of sound framings of ADMS in sociotechnical and economic terms – also has long-term advantages: as software grows ever more complex, stakeholders (including software developers and regulators) must be able to discuss ADMS, confident that their exchanges are grounded in mutual understanding (Hill, 2016).

2. Provide guidance on how to resolve tensions

   When designing and operating ADMS, tensions may arise between different ethical principles for which there are no fixed solutions (AI HLEG, 2019). Whilst EBA can help ensure compliance with a given policy, how to prioritise between conflicting interpretations of ethical principles remains a political question (Goodman, 2016). Organisations are, and should be, free to strike justifiable ethical tradeoffs within the limits of legal permissibility and operational viability. However, organisations that develop ADMS responds to a wide range of different stakeholders who often have divergent sets of interests. Regulators can facilitate the efforts of organisations to understand and account for these diverse sets of interests, e.g. by providing guidance on how to resolve tensions between conflicting values such as accuracy and privacy (Whittlestone et al., 2019), as well as on how to prioritise between conflicting definitions of normative concepts, like fairness (Kusner et al., 2017), in different situations.

3. Support the creation of standardised evaluation metrics and reporting formats

   Standardised formats for evaluation and communication help organisations and consumers to assess and compare different ADMS (Accenture, 2018). Using standardised forms of evaluation during EBA has also been shown to increase the adherence of ADMS to predefined ethics principles (Keyes et al., 2019). While organisations should be free to create or commit to different EBA procedures, regulators can strengthen the synergies between efforts by supporting the creation of standardised evaluation metrics and reporting formats. A recent example on a promising step in this direction is the ICO's (2020) *AI audit framework*.

4. Facilitate knowledge sharing and communication of best practices

   As organisations rise to the challenge of creating and implementing EBA procedures, much knowledge will be accumulated locally. It would thus be beneficial if organisations collaborated with a commitment to reproducibility, openness and the sharing of knowledge and technical solutions (Morley et al., 2020). For example, the sharing of past systems' failures could help mitigate future harms (Brundage et al., 2020). Similarly, the possibility to benchmark the performance of a system under development against that of other models would help organisations validate their ADMS with regard to relevant ethical qualities prior to its deployment (Epstein et al., 2018). To help realise the potential synergies from increased collaboration, regulators (or other government agencies) could not only provide digital platforms where software code and data could be shared but also create forums where stakeholders could discuss and share best practices for EBA of ADMS. The proposed European regulation on AI takes a step in this direction by highlighting the need to establish trusted mechanisms for the sharing and pooling of data that are essential for developing ADMS of high quality (European Commission, 2021). In the current proposal, however, the sharing of information on incidents and malfunctioning is restricted to so-called high-risk systems. To support the effectiveness of soft yet formal governance mechanisms like EBA, trusted (cross disciplinary) platforms are also needed where organisations can share information and best practices concerning all types of ADMS on a voluntary basis.

5. Create an independent body to oversee EBA of ADMS

   A plurality of approaches to EBA should be encouraged. This includes allowing actors to promote a diverse range of auditing tools and methodologies (Pedreschi et al., 2018).

Nevertheless, a specific EBA procedure will only be as good as the institution backing it (Boddington, 2017). Hence, policymakers are advised to resist the shift of power and ultimate responsibility for law enforcement from juridical courts to private actors. The solution here is to create an independent body that authorises organisations who, in turn, conduct EBA of, or issue *ethics-based* certifications for, ADMS. The primary role of such an agency would not necessarily be to evaluate and approve the design of ADMS directly (as suggested by, e.g., Tutt (2016)) but rather to inspect, verify and approve the EBA processes themselves. Given that the proposed EU regulation already sketches the contours of an EU-wide AI auditing ecosystem, one opportunity would be to leverage the same institutional structure[21] to provide assurance also for post-compliance, ethics-based audits. However, this is not the only option, and the model proposed here is not new. Within the insurance industry, for example, the European Insurance and Occupational Pensions Authority (EIOPA) works according to a similar logic with the aim of improving consumer protection, building systemic trust and enhancing supervisory convergence (EIOPA, 2018). An alternative to the creation of a new body would thus be to extend the mandate of existing, industry-specific, bodies.

6. Create incentives for voluntary adoption of EBA

Implementing EBA across organisations inevitably imposes costs, financial and otherwise (Brundage et al., 2020). Even if the aggregated benefits of EBA are justifiable in comparison to the costs, implementation will be slow if the organisations that design and deploy ADMS perceive that, for them, the costs outweigh the benefits. To incentivise the voluntary adoption of EBA, regulators could encourage and reward demonstrable achievements (Floridi et al., 2018). While such rewards could include monetary incentives like tax breaks, they could also include immaterial acknowledgements. For example, government agencies could help publish the results from EBA. They could also publish a list of organisations that adhere to specific standards or best practices.

7. Promote trust through transparency and accountability

To promote a culture of trust and collaboration, the general approach should be for businesses and regulators to work together against the problem of unethical uses of ADMS. For this reason, it is important to distinguish accountability from blame (Chopra & Singh, 2018). At the same time, people feel most obliged to obey rules and norms when these are applied fairly and consistently and when agents who deliberately break the rules see a proportionate response

---

[21] Member States will have to designate a competent national authority to supervise the application and implementation of the AIA. Importantly, this national supervisory authority should not conduct any conformity assessments itself. Instead, it will act as a notifying authority (AIA: Article 59) that assesses, designates, and notifies third-party organisations that, in turn, conduct conformity assessments of providers of high-risk AI systems.

(Hodges, 2015). Consequently, regulators could strengthen trust in emerging EBA procedures by ensuring accountability, e.g. by imposing sanctions where trust is breached. In practice, organisations that publish inaccurate, incomplete or misleading information about their ADBS could be fined or potentially lose their license to operate in a specific sector. Note that such sanctions are compatible with the voluntary nature of EBA itself. A parallel can be made to European food-labelling regulation: while food can contain both non-vegetarian and vegetarian ingredients, the malpractice of mislabelling one for the other is not permitted (Regulation (EU) No 1168, 2011).

8. Provide governmental leadership

The implementation of new governance mechanisms is most effective when part of a multi-pronged programme with executive-level support (Floridi & Strait, 2020). Similarly, the behaviour of senior management is a factor that strongly determines the chances of success of attempts to drive organisational change (Chen, 2001). The same holds on a societal level. Therefore, political leaders could help strengthen the feasibility and effectiveness of EBA as a governance mechanism by explaining and endorsing it. However, what leaders do is often more important than what they say. Hence, to demonstrate their commitment to officially stated policies, governments could consider conducting EBA of ADMS employed in the public sector and include ethics-based criteria in the public procurement of ADMS.

# 7 Concluding remarks

As important decisions that impact human lives and livelihoods are increasingly being outsourced to ADMS, it is both natural and important that these systems – as well as the organisations that design and deploy them – are subject to increased scrutiny. That said, ethical tensions do not emerge from the use of ADMS alone but can also be intrinsic to the decision-making task (Danks & London, 2017b). Moreover, both human decision-makers and ADMS come with their own sets of strengths and weaknesses.[22] Hence, overly simplistic measures – such as banning the use of ADMS for a particular task or holding ADMS to too idealistic standards that can rarely be met in full – are often unhelpful.[23] A more fruitful approach would be to understand ADMS in their specific contexts, compare them to the relative affordances and limitations of human decision-makers, and subject them to appropriate and proportional quality assurance and transparency obligations. Against this backdrop, EBA should be understood as one governance mechanism that, by contributing to procedural regularity and transparency, can help organisations – not to solve, but – to continuously manage some of the ethical risks associated with the use of ADMS.[24]

---

[22] Delegating decision-making tasks to ADMS can, for example, help reduce the level of 'noise' (Kahneman, 2021) that is characteristic of decisions made by humans – even those made by experts with access to relevant information.
[23] Issues related to 'fairness', for example, generally do not arise in a vacuum but rather as part of larger STS in which ADMS are deployed (Kearns & Roth, 2020).

A wide range of auditing tools has already been developed to help organisations validate claims made about their ADMS. This is encouraging. Yet too often, these different auditing tools are applied in isolation, each focusing on only one or a few phases in the lifecycle of ADMS. This leaves room for improvement. Since important dynamics of ADMS (as parts of larger STS) may be lost or misunderstood if subsystems are targeted separately, no purely technical solution will be able to ensure that ADMS behave in ways that are ethically-sound. However, as we have argued throughout this article, a holistic and process-oriented approach to EBA can help close this gap. The key success factor in developing feasible and effective EBA procedures is to combine existing conceptual frameworks and software tools into a structured process that monitors each phase of the ADMS lifecycle to identify and correct the points at which ethical failures (may) occur.

Understood as a *soft* yet *formal* governance mechanism, our analysis suggests that EBA can serve two main purposes. When conducted by an independent auditor and where the results are publicly communicated, EBA can provide regulators and decision-subjects with justifications for the outputs produced by ADMS. But EBA procedures can also be useful for organisations that design or deploy ADMS. For example, by informing ethical deliberation at critical intervention points that shape ADMS and their behaviour, EBA can help such organisations improve their products and services while managing reputational and financial risk. Policymakers and technology providers are thus advised to considered EBA as an integral component of multifaced approaches to ensure that ADMS are designed and deployed in ways that are ethically-sound.

At an organisational level, EBA does not necessarily require any additional layers of governance to be imposed top-down. Rather, what EBA should do is inform, formalise, assess, and interlink existing governance structures, thereby complementing and enhancing them. As discussed in section 4, that potential intervention points for EBA are therefore not limited to the engineering process of designing ADMS but extends to non-technical governance concerns such as employee training and reward structures. A good starting point would thus be for organisations to include ethics-based evaluations of ADMS in the periodic audits that they conventionally perform and complement such audits with continuous monitoring of system outputs to account for the adaptive and self-learning capabilities of ADMS. As a next step, best practice would demand that the results of such audits are communicated proactively and transparently to different relevant audiences (both internal and external).[25]

---

[24] This approach may seem radical, but is in fact not: research ethics and medical ethics have always involved a combination of the law, ethical governance policies, practices, and procedures, with contextual discursive and procedural support (Morley et al., 2021).

While private organisations – both providers of ADMS and professional service firms – have good reasons to develop and voluntary implement EBA procedures, there is much policymakers can do to support the feasibility and effectiveness of such initiatives. To recapitulate our findings from section 6, policymakers and regulators should consider: supporting the creation of standardised evaluation metrics and reporting formats; providing working definitions for ADMS; providing guidance on how to resolve tensions; facilitating knowledge sharing; creating an independent body to oversee EBA; creating incentives for voluntary adoption of EBA procedures, promoting trust through transparency and accountability, and providing governmental leadership. All these policy recommendations are based on the general approach that businesses and regulators can and should work together against the problem of unethical uses of ADMS.[26]

That said, let us conclude with a word of caution. As stressed throughout this article, by contributing to procedural regularity and transparency, EBA can support public reasoning about what constitutes ethical decision-making in different contexts and inform ethical deliberation within organisations that design and deploy ADMS. However, EBA does not make existing governance structures redundant, nor should it be used as an excuse to delay necessary legislation. Rather, EBA should be viewed as a complement to existing governance mechanisms like codes of conduct and certification. Further, as a new governance mechanism, EBA is inevitably exposed to the risks and limitations discussed in section 5. To reiterate, these risks include ethics shopping, ethics bluewashing and ethics lobbying, on the one hand, and tunnel vision, random agenda selection and inconsistent policymaking, on the other. To navigate these risks, regulators and businesses are advised to pilot different EBA procedures in controlled settings and allow for an iterative process to illuminate and assist the wider adoption of such practices.

To conclude, EBA will not, and should not, replace the need for continuous ethical reflection amongst designers, operators, and users of ADMS. Nevertheless, if implemented according to the policy recommendations listed above, EBA can help society manage some of the ethical risks posed by ADMS while reaping their economic and social benefits.

---

[25] The concept of transparency has multiple uses in various disciplines. Larsson & Heintz (2020) argue that transparency in AI is best understood as a balancing of interests that require an understanding of local contexts and information asymmetries. The key takeaway for our purposes is that EBA should focus on local interpretability, i.e. explanations targeted at specific stakeholders – such as decision subjects or external auditors – and for specific purposes like internal governance, external reputation management, or third-party verification.

[26] As noted by Floridi (2014b), business, at its core, is an ethically good force when directed towards supplying goods and services whilst reducing waste.